\documentclass[11pt]{article}
\usepackage{graphicx}

\newcommand{\BABARPubYear}    {04}

\newcommand{\BABARConfNumber} {027}
\newcommand{\SLACPubNumber} {10631}

\input pubboard/babarsym

\def\DsTT{D_{sJ}^*(2317)^+}
\def\DsTO{D_s^{*}(2112)^+}
\def\DsFE{D_{sJ}(2460)^+}
\def\DsTS{D_{s1}(2536)^+}

\def\DsTTdc{D_{sJ}^*(2317)^{++}}
\def\DsTTz{D_{sJ}^*(2317)^0}

\long\def\inst#1{\par\nobreak\kern 4pt\nobreak
    {\it #1}\par\vskip 10pt plus 3pt minus 3pt}

\begin{document}
{\pagestyle{empty}

\begin{flushright}
\babar-CONF-\BABARPubYear/\BABARConfNumber \\
%\babar-PUB-\BABARPubYear/\BABARPubNumber \\
SLAC-PUB-\SLACPubNumber \\
%hep-ex/\LANLNumber \\
August, 2004 \\
\end{flushright}

\par\vskip 5cm

% Title of the paper
\begin{center}
\Large \bf {\boldmath Measurement of the $\DsTT$ and $\DsFE$
Properties
in $e^+ e^- \to c \overline{c}$ Production}
\end{center}
\bigskip

\begin{center}
\large The \babar\ Collaboration\\
\mbox{ }\\
\today
\end{center}
\bigskip \bigskip

% Abstract 
\begin{center} 
\large \bf Abstract 
\end{center}
The properties of the
$\DsTT$ and $\DsFE$ mesons are studied using 125~${\rm
fb}^{-1}$ of $e^+e^- \to c \overline{c}$ data collected by the BaBar
experiment. Preliminary mass estimates of
$[2318.9\pm 0.3\;({\rm stat.}) 
          \pm 0.9\;({\rm syst.})]$~\mevcc and
$[2459.4 \pm 0.3\;({\rm stat.}) 
          \pm 1.0\;({\rm syst.})]$~\mevcc
are obtained. Searches are performed for the
decay to the $D_s^+$ meson along with one or more $\piz$, $\pi^+$, or $\gamma$
particles. A search is also performed for neutral and doubly-charged partners.

\vfill
\begin{center}

Submitted to the 32$^{\rm nd}$ International Conference on High-Energy Physics, ICHEP 04,\\
16 August---22 August 2004, Beijing, China

\end{center}

\vspace{1.0cm}
\begin{center}
{\em Stanford Linear Accelerator Center, Stanford University, 
Stanford, CA 94309} \\ \vspace{0.1cm}\hrule\vspace{0.1cm}
Work supported in part by Department of Energy contract DE-AC03-76SF00515.
\end{center}

\newpage
} % end of pagestyle{empty}

% Input author list file
%
%author list removed temporarily to save trees 7/9/04 RNC
%
\begin{center}
\small

The \babar\ Collaboration,
\bigskip

%% author list as of 02-Jul-2004 (609 authors)
%
B.~Aubert,
R.~Barate,
D.~Boutigny,
F.~Couderc,
J.-M.~Gaillard,
A.~Hicheur,
Y.~Karyotakis,
J.~P.~Lees,
V.~Tisserand,
A.~Zghiche
\inst{Laboratoire de Physique des Particules, F-74941 Annecy-le-Vieux, France }
% ...

\end{center}\newpage

% The body of the paper starts here
\section{INTRODUCTION}
\label{sec:Introduction}

The $\DsTT$ meson, discovered by this collaboration~\cite{Aubert:2003fg}
and confirmed by others~\cite{Besson:2003cp,Abe:2003jk,Krokovny:2003zq}, and 
the $\DsFE$ meson,
observed by the CLEO collaboration~\cite{Besson:2003cp} and 
confirmed by this collaboration~\cite{Aubert:2003pe}
and others~\cite{Abe:2003jk}, 
has reawakened interest in the study of the spectroscopy of
charm mesons.

Presented in this paper is an updated, preliminary analysis of these 
states using 125~${\rm
fb}^{-1}$ of $e^+e^- \to c \overline{c}$ data collected by the \babar\ 
experiment. From this analysis we present new estimates of the
$\DsTT$ and $\DsFE$ masses and the branching ratios of $\DsFE$
decays to $\Ds\gamma$ and $\Ds\pip\pim$ with respect to
its decay to $\Ds\piz\gamma$. In addition, we search for new decays
of either meson involving
a $\Ds$ meson accompanied by $\piz$ and $\pi^\pm$ mesons and photons.
An estimate of the $\DsTS$ mass is also presented.

These measurements are performed by fitting the invariant mass spectrums
of combinations~\footnote{Inclusion of charge conjugate states is implied 
throughout this paper.} of $\Ds\piz$, 
$\Ds\gamma$, $\Ds\piz\gamma$, and
$\Ds\pip\pim$ particles. Combinations of $\Ds\pip$ and $\Ds\pim$ are also
studied in the search for new states. The following two sections of this
paper describe those details in common to the entire analysis.
The study of each combination is then discussed in their own section. 
The paper ends with a summary.

\section{CANDIDATE SELECTION}
\label{sec:babar}

This analysis is performed using a 125~${\rm fb}^{-1}$
data sample collected on or near the $\Upsilon(4S)$ resonance
with the \babar\  detector at the
\pep2 asymmetric-energy $e^+e^-$ storage rings.
The \babar\ detector,
a general-purpose, solenoidal, magnetic spectrometer, is
described in detail elsewhere \cite{Aubert:2001tu}. 
Charged particles were detected
and their momenta measured by a combination of a drift chamber (DCH)
and silicon vertex tracker (SVT), both operating within a
1.5-T solenoidal magnetic field. 
A ring-imaging Cherenkov detector (DIRC) is used for
charged-particle identification. Photons are detected and
measured with a CsI electromagnetic calorimeter (EMC).

All of the final states explored in this analysis involve one
$\Ds$ candidate decaying to $K^+K^-\pi^+$. 
A clean sample of $K^\pm$ candidates is obtained using particle
identification by requiring a Cherenkov photon yield and angle
consistent with the $K^\pm$ hypothesis. This information
is augmented with energy loss measurements in the tracking systems.
The efficiency of $K^\pm$ identification is approximately 85\% in the
kinematic range used in this analysis with a $\pi^\pm$ misidentification
rate
of less than 2\%. A similar procedure is used to produce a sample
of $\pi^\pm$ candidates.

Each $\Ds$ candidate is constructed by
combining a $\Kp \Km$ candidate pair with a $\pip$ candidate 
in a geometrical
fit to a common vertex. An acceptable $\Kp\Km\pip$ candidate must have a fit
probability greater than 0.1\% and a trajectory consistent with 
originating from the $e^+e^-$ 
luminous region.
Backgrounds are further suppressed by selecting 
decays to $\Kbar^{*0}K^+$ and $\phi\pip$.
Additional details of this selection procedure
can be found elsewhere~\cite{Aubert:2003fg}.  
Combinations of $K^-K^+\pi^+$
with $1.954<m(\Kp\Km\pip)<1.981$~\gevcc are taken as $\Ds$ candidates.

For the purposes of calculating the invariant mass of the various
particle combinations in this paper,
the energy of each $\Ds$ candidate is calculated from the measured
momenta and the PDG $\Ds$ mass of $1968.5$~\mevcc~\cite{Hagiwara:2002pw}.
The uncertainty in this mass ($0.6$~\mevcc)
is taken as a systematic uncertainty.

Once a $\Ds$ candidate is obtained, a search is performed for
all accompanying $\pi^0$, $\gamma$, and $\pi^\pm$ particles.
All energy clusters in the EMC unassociated with charged tracks and
consistent with an electromagnetic shower are considered photon candidates.
A candidate $\piz$ is formed by constraining a pair 
of photons each with energy greater than 100~MeV
to emanate from
the intersection of the $\Ds$ trajectory
with the beam envelope, 
performing a one-constraint fit to the
$\piz$ mass, and requiring a fit probability 
greater than 2\%.  

A given event
may yield several acceptable $\piz$ candidates. For the various
final-state samples used in this analysis, we retain only
those candidates for which neither 
photon belongs to another otherwise acceptable $\piz$ of momentum
greater than 150~\mevc. In a similar fashion, we discard any
$\gamma$ candidate which belongs to the same list of
acceptable $\piz$ candidates.

\section{SIMULATION AND CALIBRATION}

The efficiency and reconstructed mass distribution of all
final-state combinations discussed in this document are derived
from detailed simulation based on the Geant4~\cite{Agostinelli:2002hh} 
toolkit,
reconstructed with the same algorithms as the data,
and passing the same selection requirements.
Individual samples of between 120,000 and 240,000 events were generated 
for each individual signal.

In addition, for the purposes of optimizing candidate selection
requirements,
a simulated sample corresponding to 30~${\rm fb}^{-1}$ of 
$e^+e^-\to c\overline{c}$ events is used to describe combinatorial
background.

Clean samples of $\tau\to\rho\nu$ and $\tau\to\pi\nu$ were studied
to validate the accuracy of the simulation
of the reconstruction efficiency and energy
scale for photon and $\piz$ candidates. Based on this study,
a photon efficiency correction of 1.6\% and EMC energy scale
corrections of between 0.2\% and 0.8\% is applied.
The remaining systematic uncertainty in the Monte Carlo
prediction of neutral efficiency and energy scale is a relative 3\% and
1\%, respectively.

The reconstruction and selection efficiency of the various final-state
combinations studied in this analysis tend to be dependent on the
center-of-mass momentum $p^*$ of the $\DsTT$ and $\DsFE$.
To minimize any dependence on the $p^*$ distribution assumed
in the Monte Carlo, each final-state is restricted to the same
minimum $p^*$ value of 3.2~\gevcc (this also serves to remove production
from $B$ decay). In addition, a $p^*$ dependent
efficiency function derived from Monte Carlo
is used to calculate efficiency corrected yields for all decay modes.

Estimates of the tracking efficiency
obtained from $\tau$ decays and from a comparison of SVT and DCH
measurements indicate that the Monte Carlo prediction of $\pi^\pm$
efficiency is, on average, 0.8\% too high, with a systematic uncertainty 
of 1.3\%.

\section{\boldmath THE $\Ds\piz$ SYSTEM}
\label{sec:dspi0}

\begin{figure}
\begin{center}
\includegraphics[width=0.7\linewidth]{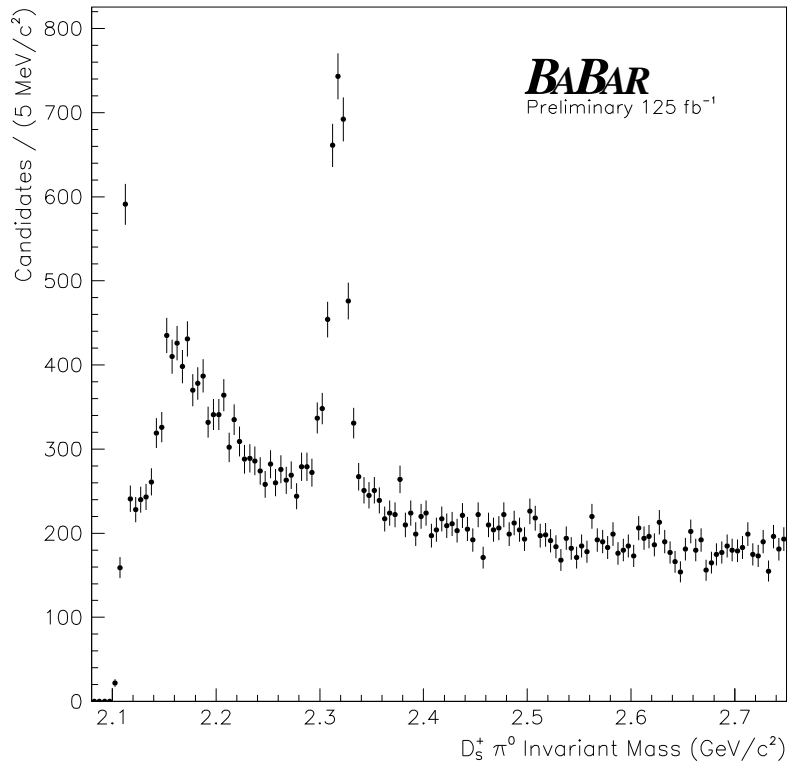}
\end{center}
\vskip -0.35in
\caption{\label{fg:dspi0.nocut}The $\Ds\piz$ invariant
mass distribution for candidates with loose $\piz$ requirements.
}
\end{figure}

To form $\Ds\piz$ combinations, each $\Ds$ candidate is combined
with one $\piz$ candidate. The $p^*$ of the resulting combination
is required to be greater than 3.2~\gevc.
The corresponding invariant mass distribution
is shown in Fig.~\ref{fg:dspi0.nocut}. Signals for $\DsTT$ and $\DsTO$
are clearly visible. 

\begin{figure}
\begin{center}
\includegraphics[width=0.7\linewidth]{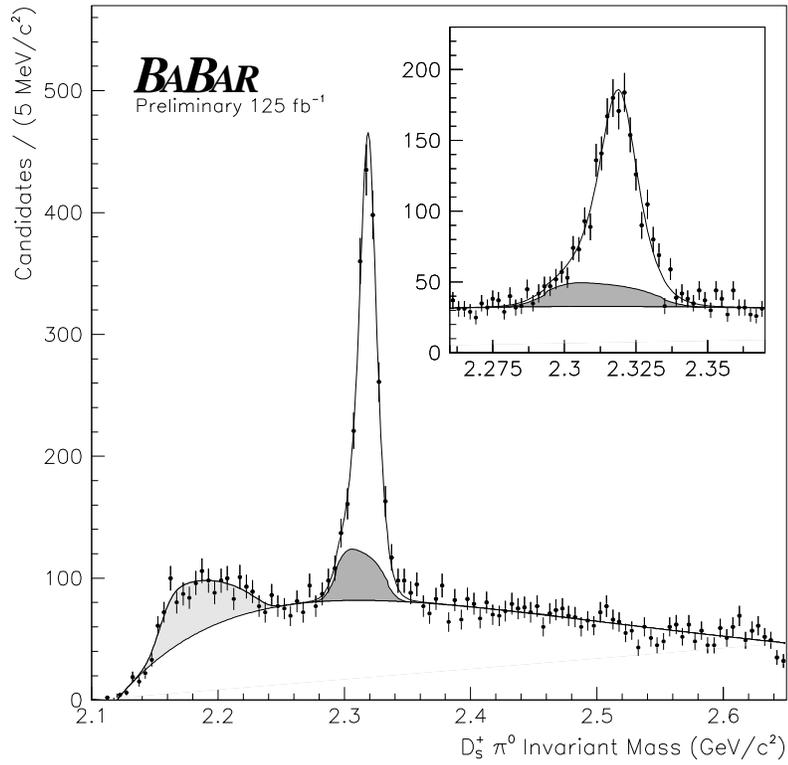}
\end{center}
\vskip -0.35in
\caption{\label{fg:dspi0.fit}The $\Ds\piz$ invariant
mass distribution for candidates that satisfy the requirements
discussed in the text. The solid curve is the result of a unbinned
likelihood fit. The dark (light) gray region is the contribution from the
$\DsFE\to\Ds\piz\gamma$ ($\DsTO\to\Ds\gamma$) reflection.
The inset is an expanded view near the $\DsTT$ mass.
}
\end{figure}

For the purposes of fitting the $\Ds\piz$
mass spectrum and extracting information on the $\DsTT$ and
$\DsFE$ mesons, the $\Ds\piz$ sample is restricted to
those $\piz$ candidates with momentum greater than 400~\mevc.
This momentum requirement is based on an optimization of
$Q = \varepsilon/\sqrt{B}$, where $\varepsilon$ is the efficiency
for the hypothetical decay $\DsFE\to\Ds\piz$,
and $B$ is the amount of background as predicted by Monte Carlo.
The resulting invariant mass distribution is shown in
Fig.~\ref{fg:dspi0.fit}. As a result of this stricter $\piz$ momentum
requirement, the $\DsTO\to\Ds\piz$ signal is entirely eliminated.

Estimates of the yield and mass of the $\DsTT$ are obtained 
using a unbinned likelihood fit of the $\Ds\piz$ mass spectrum.
The lineshape is also used to calculate a limit on the yield from 
hypothetical $\DsFE\to\Ds\piz$ decays.
The lineshapes of the reconstructed $\DsTT$ and $\DsFE$ mass distributions
are obtained from fits to simulated signal events. In both the
simulation and in the fit it is assumed that the
intrinsic width $\Gamma$ of both mesons is small enough that
this lineshape is not significantly affected.

\begin{figure}
\begin{center}
\includegraphics[width=0.7\linewidth]{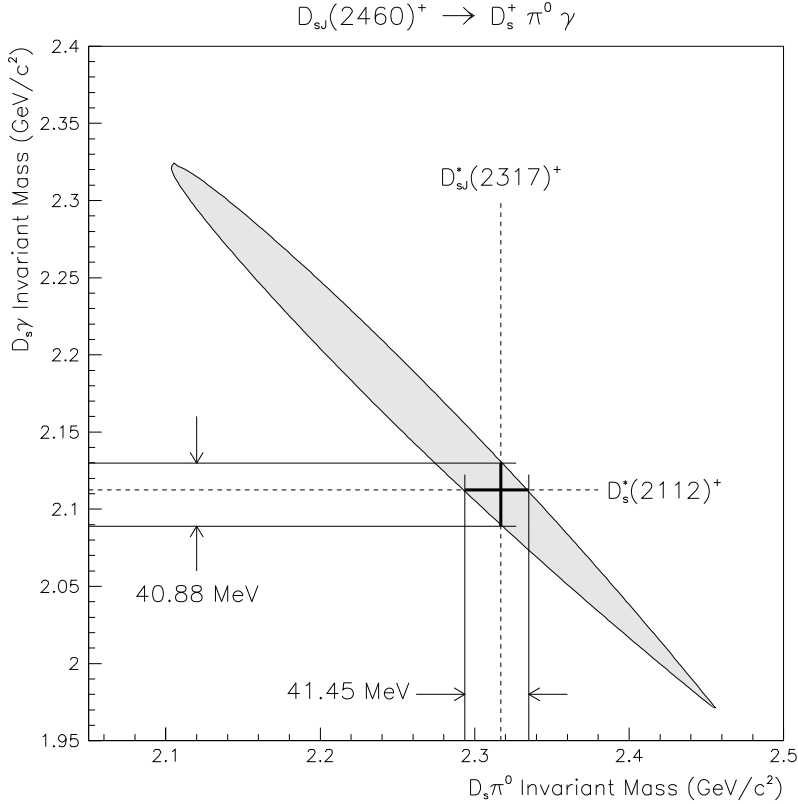}
\end{center}
\vskip -0.35in
\caption{\label{fg:dspi0gam.dalitz}The kinematics of $\DsFE\to\Ds\piz\gamma$
decay. The shaded area indicates the region of $\Ds\piz$ and $\Ds\gamma$
mass that is kinematically allowed in the decay of an object of mass
$2459.2$~\mevcc. The lines mark the kinematic space limited to the
decay which proceeds through intermediate $\DsTO$ or $\DsTT$ mesons. 
}
\end{figure}

The background to the $\DsTT$ comes from unrelated $\Ds$ and
$\piz$ (the combinatorial background) and from two types of
reflections. One reflection, from $\DsTO\to\Ds\gamma$ decays
combined with an unassociated $\gamma$ to form a fake $\piz$ candidate,
is peaked near the kinematic limit in $\Ds\piz$ mass of 2154.6~\mevcc.
The second reflection is produced by the $\Ds$ and $\piz$
mesons from $\DsFE\to\DsTO\piz$ decay. Due to a kinematic
coincidence, this reflection has a mean mass that is close
to the $\DsTT$ mass and must be accurately determined in order
to correctly measure the $\DsTT$ properties.
As shown in Fig.~\ref{fg:dspi0gam.dalitz}, if reconstruction
efficiency is uniform, the $\DsFE$ reflection would 
appear as a nearly square function of total width $41.5$~\mevcc
smeared by resolution. Because of the variation in reconstruction
efficiency across the $\DsFE$ phase space, 
the reflection shape is asymmetric. The prediction from $\DsFE$ Monte
Carlo is used to determine the precise shape.

The likelihood fit to the $\Ds\piz$ mass spectrum is shown
in Fig.~\ref{fg:dspi0.fit}. Whereas the shape and magnitude 
of the $\DsFE$ reflection are
fixed to Monte Carlo predictions, the parameters of the $\DsTO$ reflection
and the shape of the combinatorial background are allowed to vary 
in the fit. The mass and yield of the $\DsTT$ are determined by
shifting the associated lineshape up or down in mass and adjusting the
overall amplitude to best match the data. A limit on the yield
of the hypothetical $\DsFE\to\Ds\piz$ decay is determined by adjusting 
the amplitude of the associated lineshape positive or negative to best
match the data. The resulting likelihood fit successfully describes
the data.

The fit determines (statistical errors only)
a $\DsTT$ mass of [$2318.9\pm 0.3$]~\mevcc and 
raw $\DsTT$ and $\DsFE$ yields of $1275 \pm 45$ and $3 \pm 26$ mesons.

A systematic uncertainty specific to the fit to the $\Ds\piz$ system is 
the size of the $\DsFE$ reflection. If the likelihood fit is allowed to
adjust the size of this reflection to best match the data, the result is
a $\DsTT$ yield 0.5\% smaller and a shift in $\DsTT$ mass that is less
than 0.1~\mevcc. In addition, various different models of the $\DsTO$
reflection can be used in the fit, in which case variations in the
$\DsTT$ yield of up to 0.4\% are observed.

\section{\boldmath THE $\Ds\gamma$ SYSTEM}
\label{sec:dsgam}

\begin{figure}
\begin{center}
\includegraphics[width=0.7\linewidth]{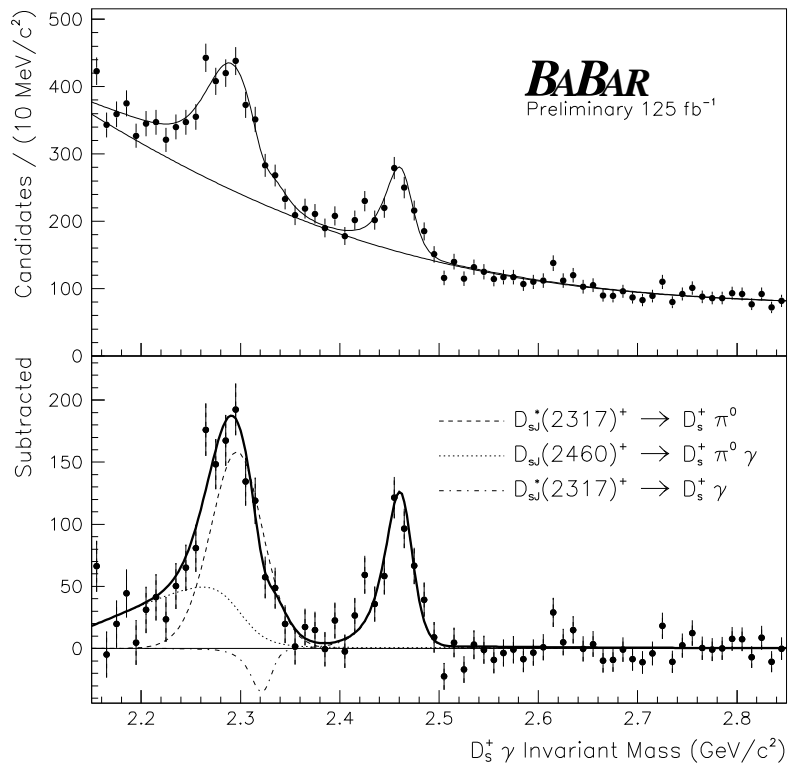}
\end{center}
\vskip -0.35in
\caption{\label{fg:dsgam.fit} (top) The $\Ds\gamma$ invariant
mass distribution for candidates that satisfy the requirements
discussed in the text. The solid curve is the result of a unbinned
likelihood fit. (bottom) The mass distribution after subtracting
the contribution from combinatorial background as estimated by the fit.
Various contributions to the signal and reflection portions of the
fit are overlaid. The $\DsTT$ signal (dot-dash line) appears as a negative
contribution due to a negative fluctuation in the fit.
}
\end{figure}

To form $\Ds\gamma$ combinations, each $\Ds$ candidate is combined
with $\gamma$ candidates with energy greater than 500~\mevc.
This energy requirement is based on an optimization of
$Q' = S/\sqrt{S+B}$, where $S$ is the expected signal size~\cite{Abe:2003jk}
and $B$ is the amount of background as predicted by Monte Carlo
for the decay $\DsFE\to\Ds\gamma$. In addition, any
$\Ds\gamma$ combination with $p^*$ less than 3.2~\gevc is discarded.
The resulting invariant mass distribution is shown in
Fig.~\ref{fg:dsgam.fit}.

For reasons of simplicity, the likelihood fit is restricted to
a $\Ds\gamma$ mass range between 2.15 and 2.85~\mevcc. As shown
in Fig.~\ref{fg:dsgam.fit}, there
are two structures clearly visible in this spectrum within
this range on top of a gradually falling background distribution.
The higher mass structure corresponds to 
the $\DsFE$ meson. The likelihood fit
uses a signal lineshape determined from signal Monte Carlo.
To determine the $\DsFE$ mass, the mean value of the signal is allowed
to shift. To determine the $\DsFE$
yield, the fit determines the amplitude that bests describes the data.

The lower mass structure is a combination of two reflections. The largest
reflection is composed of 
the $\Ds$ meson from the decay $\DsTT\to\Ds\piz$ combined
with one of the photons from the $\piz$. The second
is produced in a similar fashion from $\DsFE\to\Ds\piz\gamma$ decay.
The shapes of these two reflections are
influenced
by the candidate selection requirements and, in particular, the
requirement that the $\gamma$ particle not belong to a $\piz$
candidate. This interdependency is complex enough that the Monte Carlo
simulation fails to adequately reproduce the precise shape of the
reflection. For this reason a parameterization of $\DsTT\to\Ds\piz$
reflection shape is included as part of the fit. The $\DsFE\to\Ds\piz\gamma$
reflection shape is fixed to the prediction from Monte Carlo.

The kinematics of $\DsTT\to\Ds\piz$ decay require that the
$\Ds\gamma$ reflection end at $\Ds\gamma$ mass of 2.304~\mevcc.
The lineshape assumed by the likelihood fit respects this limit
but includes detector resolution effects that will tend to blur
this boundary. Just adjacent to the boundary is where a hypothetical
$\DsTT\to\Ds\gamma$ signal would reside. The lineshape of the $\DsTT$
signal is extracted from signal Monte Carlo. Because of uncertainties
in the precise shape of the $\DsTT\to\Ds\piz$ and
$\DsFE\to\Ds\piz\gamma$ reflections, there
is considerable uncertainty in the potential amount of
$\DsTT\to\Ds\gamma$ decays.

The likelihood fit represents the background by a constant plus
the tail of a Gaussian distribution. The result of the fit is shown
in Fig.~\ref{fg:dsgam.fit}. A $\DsFE$ mass of [$2457.2\pm 1.6$]~\mevcc
and yield of $509\pm 46$ mesons is obtained (statistical errors only). 
The fit, which allows the signal yield
to fluctuate to negative values, obtains $-107\pm 84$ $\DsTT$ decays
(statistical errors only).

The largest uncertainty in the likelihood fit is the shape
of the reflection and background. Any tails in the signal
or reflection mass distributions will tend to fill in the region of mass 
around 2.35~\mevcc,
suppressing the size of the background and inflating signal yields.
In addition, as mentioned above, the $\DsTT$ yield is sensitive to the shape
of the reflections because the two peaks overlap. 
In particular, if the Monte Carlo predictions
for the reflect shape and size is used in the fit, the $\DsTT$ yield
is increased to $209$ mesons. 

Another systematic check is to use alternate functions to describe
the combinatorial background, in which case it is possible to change
the $\DsFE$ yield by at most 3\% without adversely affecting the
fit quality.

\section{\boldmath THE $\Ds\piz\gamma$ SYSTEM}
\label{sec:dspi0gam}

A likelihood fit to the
$\Ds\piz\gamma$ invariant mass distribution is used to determine
the mass and yield of the $\DsFE$ meson. Because this is a three body
decay, there are additional issues that need to be addressed.
In particular, it is reasonable to assume that the $\DsFE$ decays
to $\Ds\piz\gamma$ through either of two possible intermediate resonances:
$\DsTO\piz$ or $\DsTT\gamma$. As shown in Fig.~\ref{fg:dspi0gam.dalitz},
these two decay modes overlap in all three mass projections.
There is no {\em a priori} reason to favor one decay path over another.
A second, two-dimensional fit is used to distinguish them.

To select the data sample for studying the $\DsFE$ meson,
the quantity $Q' = S/\sqrt{S+B}$ is optimized simultaneously for both
the minimum $\piz$ momentum and $\gamma$ energy.
The expected signal size $S$ is calculated based on
results from previous measurements~\cite{Aubert:2003pe}. Based on this
procedure a minimum $\piz$ momentum of 400~\mevc and minimum
$\gamma$ energy of 135~\mev is chosen.

\begin{figure}
\begin{center}
\includegraphics[width=0.7\linewidth]{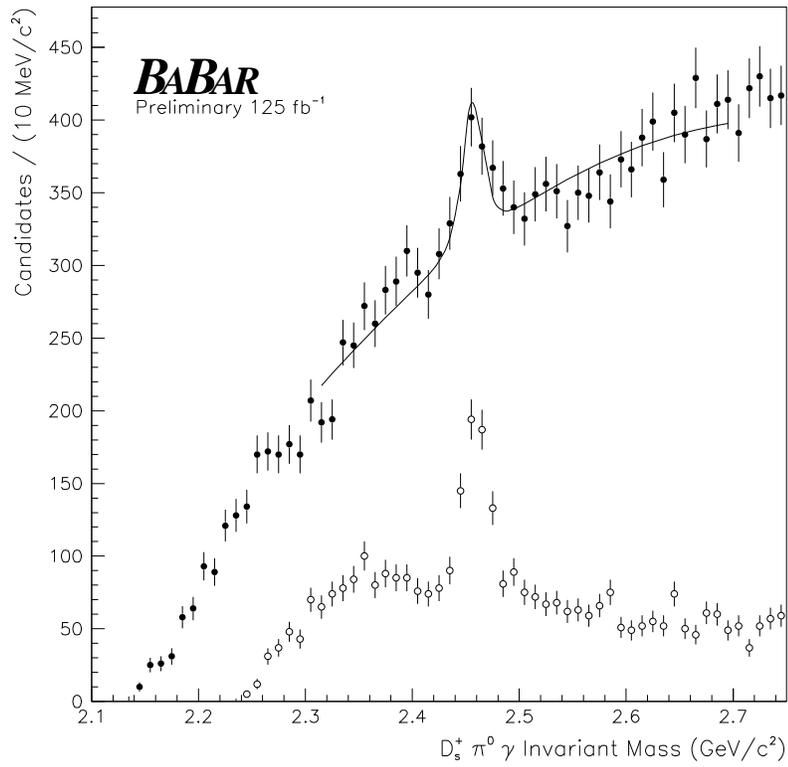}
\end{center}
\vskip -0.35in
\caption{\label{fg:dspi0gam.dataall} The inclusive and semi-inclusive
$\Ds\piz\gamma$ invariant mass distributions. Shown in solid points is the
total data sample. Overlaid is the simple line fit discussed in the text. 
Shown in open round points is the
subset which has a $\Ds\gamma$ mass within $\pm 2$~\mevcc of the $\DsTO$ mass.
}
\end{figure}

Included in Fig.~\ref{fg:dspi0gam.dataall} is the $\Ds\piz\gamma$
mass spectrum for all selected candidates below a mass of 2.75~\gevcc.
A peak near the $\DsFE$ mass is apparent. The signal shape of the
$\DsFE$ is obtained by fitting a signal Monte Carlo sample. This
shape is combined with a second order polynomial to fit the
mass spectrum. The result is a $\DsFE$ mass of [$2457.8\pm 2.8$]~\mevcc
and yield of $246 \pm 44$ mesons (statistical errors only). 
The $\DsFE$ peak, however, is underlayed by
a substantial background which has structure that is not well
represented by a simple polynomial. To obtain a more accurate measurement 
of the $\DsFE$ meson requires a more refined selection to
isolate the signal.

Since both the $\DsTO\piz$ and $\DsTT\gamma$ decay modes are
restricted to a $\Ds\gamma$ mass near the $\DsTO$ mass, the $\DsFE$
signal can be isolated by requiring the $\Ds\gamma$ invariant mass
to reside within 2~\mevcc of the $\DsTO$ mass 
($2112.4$~\mevcc~\cite{Hagiwara:2002pw}). The result is included
in Fig.~\ref{fg:dspi0gam.dataall}. Although the background is now
substantially reduced, it is clear that some type of peaking background
has been introduced by this requirement since the size of the peak in
the restricted data sample is noticeably larger than that of the inclusive
sample.

\begin{figure}
\begin{center}
\includegraphics[width=0.7\linewidth]{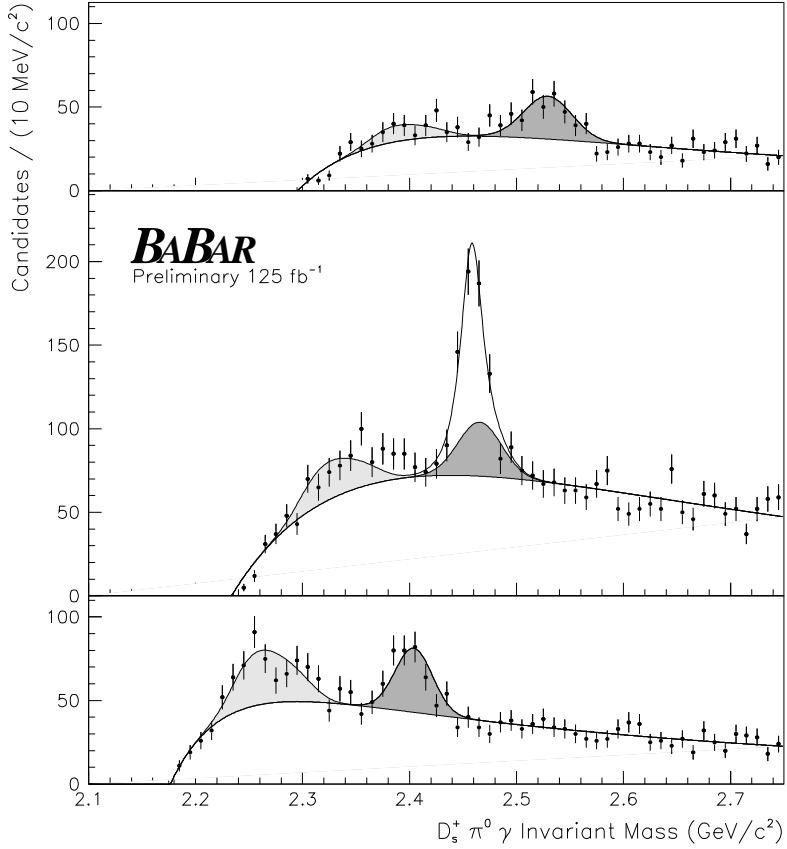}
\end{center}
\vskip -0.35in
\caption{\label{fg:dspi0gam.fit}The $\Ds\piz\gamma$ mass spectrum
of the sample of events that fall in a $\Ds\gamma$ signal region (center
plot), the $\Ds\gamma$ high mass sideband (top plot), and the
low mass sideband (bottom plot). The $\DsFE$
signal appears only in the signal region. The results of separate
likelihood fits are superimposed on all three distributions.
Reflections from (dark gray) 
$\DsTT\to\Ds\piz$ and (light gray) $\DsTO\to\Ds\gamma$ decays
appear in all three distributions at positions that are fixed from
kinematics. The relative contribution of the $\DsTT$ reflection 
in each sample has been determined using signal Monte Carlo.
}
\end{figure}

The source of the peaking background can be investigated further
by considering two $\DsTO$ sidebands, each of total width
$40$~\mevcc in $\Ds\gamma$ mass and with centers separated 
by $\pm 60$~\mevcc from
the $\DsTO$ mass. The $\Ds\piz\gamma$ mass distribution
in these sidebands is included in Fig.~\ref{fg:dspi0gam.fit}.
Two reflections are observed. Both can be identified in the
$\Ds\piz$ invariant mass distribution of Fig.~\ref{fg:dspi0.fit}.
The peak that appears near a $\Ds\piz\gamma$ mass of 2.25~\mevcc
in the lower $\Ds\gamma$ sideband originates from
$\DsTO\to\Ds\gamma$ decays combined with two unassociated $\gamma$ particles,
one of which has been combined with the $\gamma$ from the $\DsTO$
decay to form a fake $\piz$ candidate. A calculation of kinematics
based only on the $\Ds$, $\piz$, and $\DsFE$ masses
can demonstrate that the $\DsTO$ reflection will appear with
a predictable amount of smearing at
a specific $\Ds\piz\gamma$ mass in the $\Ds\gamma$ signal and sideband
selections.

A more important reflection is from $\DsTT\to\Ds\piz$ decays combined
with unassociated $\gamma$ particles. The position
and lineshape of this reflection has been determined using $\DsTT$
Monte Carlo. Although the Monte Carlo prediction of the size of the
reflection is consistent with the data within statistics, its size
has been adjusted uniformly 
upwards by 23\% (corresponding to approximately one standard deviation)
to best match the amplitudes observed in the two
$\Ds\gamma$ sidebands. The accuracy of the Monte Carlo prediction
of the $\DsTT$ shape has already
been confirmed in the $\Ds\piz$ fit shown in Fig.~\ref{fg:dspi0.fit}

Independent likelihood fits are applied to each of the sideband
distributions and to the signal distribution. All three fits can
successfully model their respective data samples, as shown in 
Fig.~\ref{fg:dspi0gam.fit}. The fit to the
$\Ds\gamma$ signal region includes those two reflections and
the $\DsFE$ signal. The $\DsFE$ mass is included in the fit by allowing
the lineshape to up or down in mass. The fit obtains a
$\DsFE$ mass of [$2459.1\pm 1.3$]~\mevcc and yield of $292\pm 29$ mesons
(statistical errors only).

A systematic uncertainty specific to this fit is the size of the
$\DsTT$ reflection. Using the prediction of yield from the Monte Carlo
unchanged increases the $\DsFE$ yield to $308$ mesons and shifts the
mass upwards by 0.3~\mevcc.

The fit of the $\Ds\piz\gamma$ mass does little to differentiate
between the two possible $\DsFE$ decay modes $\DsTO\piz$ and
$\DsTT\gamma$. It is necessary for this purpose to investigate
the $\Ds\piz$ and $\Ds\gamma$ mass distributions more closely.
As shown in Fig.~\ref{fg:dspi0gam.dalitz}, the $\DsTO\piz$ decay 
can be expected to appear
as a sharp peak in the $\Ds\gamma$ spectrum centered on the $\DsTO$
mass, and with a distribution spread out by approximately
41.3~\mevcc in $\Ds\piz$ mass centered at $2313.4$~\mevcc. 
In contrast, the decay $\DsFE \to \DsTT \gamma$
appears as a sharp peak in the $\Ds\piz$ spectrum and as a distribution
spead out by approximately 40.9~\mevcc
in the $\Ds\gamma$ spectrum centered at $2110.5$~\mevcc.

\begin{figure}
\begin{center}
\includegraphics[width=0.7\linewidth]{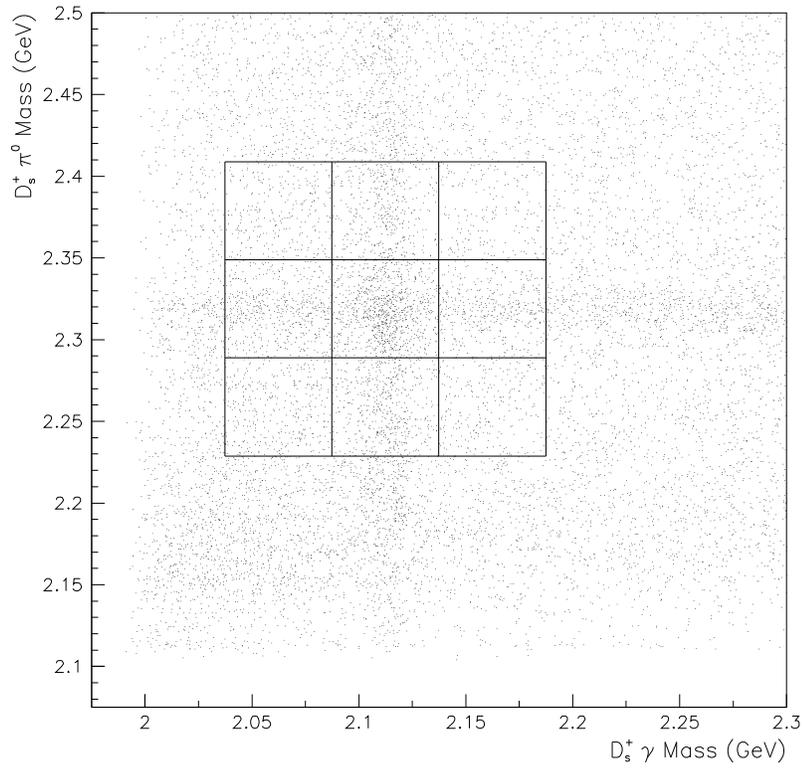}
\end{center}
\vskip -0.35in
\caption{\label{fg:dspi0gam.range}A scatter plot of the $\Ds\piz$
versus $\Ds\gamma$ invariant masses from the sample of $\Ds\piz\gamma$
candidates. Two bands from $\DsTT\to\Ds\piz$ and $\DsTO\to\Ds\gamma$
decays are clearly visible as background. 
The $\DsFE$ signal consists of an excess
of candidates near the area these two bands cross. The grid lines indicate 
the range of events used in a likelihood fit.
}
\end{figure}

The $\Ds\piz$ and $\Ds\gamma$ mass distributions of the signal
cannot be explored without correctly subtracting backgrounds
from $\DsTO\to\Ds\gamma$ and $\DsTT\to\Ds\piz$ decays.
This
subtraction is performed by a two-dimensional unbinned likelihood applied
to the $\Ds\piz$ and $\Ds\gamma$ mass distributions of the data.
The likelihood fit is restricted to the data sample contained in
the boundaries illustrated
in Fig.~\ref{fg:dspi0gam.range}. This fit includes five sources of
$\Ds\piz\gamma$ candidates:
\begin{itemize}
\item
Combinatorial background represented by a two-dimensional quadratic
function.
\item
Background from $\DsTO\to\Ds\gamma$ decays combined with unassociated
$\piz$ candidates represented by a $\DsTO$ lineshape in the $\Ds\gamma$
mass and as a linear function in $\Ds\piz$ mass.
\item
Background from $\DsTT\to\Ds\piz$ decays combined with unassociated
$\gamma$ candidates represented by a $\DsTT$ lineshape in the $\Ds\piz$
mass and as a linear function in $\Ds\gamma$ mass.
\item
A signal from $\DsFE\to\DsTO\piz$ represented by a $\DsTO$ lineshape
in the $\Ds\gamma$ mass and a $41.5$~\mevcc wide reflection
in the $\Ds\piz$ mass with a specific shape and mass predicted by
signal Monte Carlo.
\item
A signal from $\DsFE\to\DsTT\gamma$ represented by a $\DsTT$ lineshape
in the $\Ds\piz$ mass and a $40.9$~\mevcc wide reflection
in the $\Ds\gamma$ mass with a shape determined from kinematics
smeared by a resolution extracted from signal Monte Carlo.
\end{itemize}

\begin{figure}
\begin{center}
\includegraphics[width=0.8\linewidth]{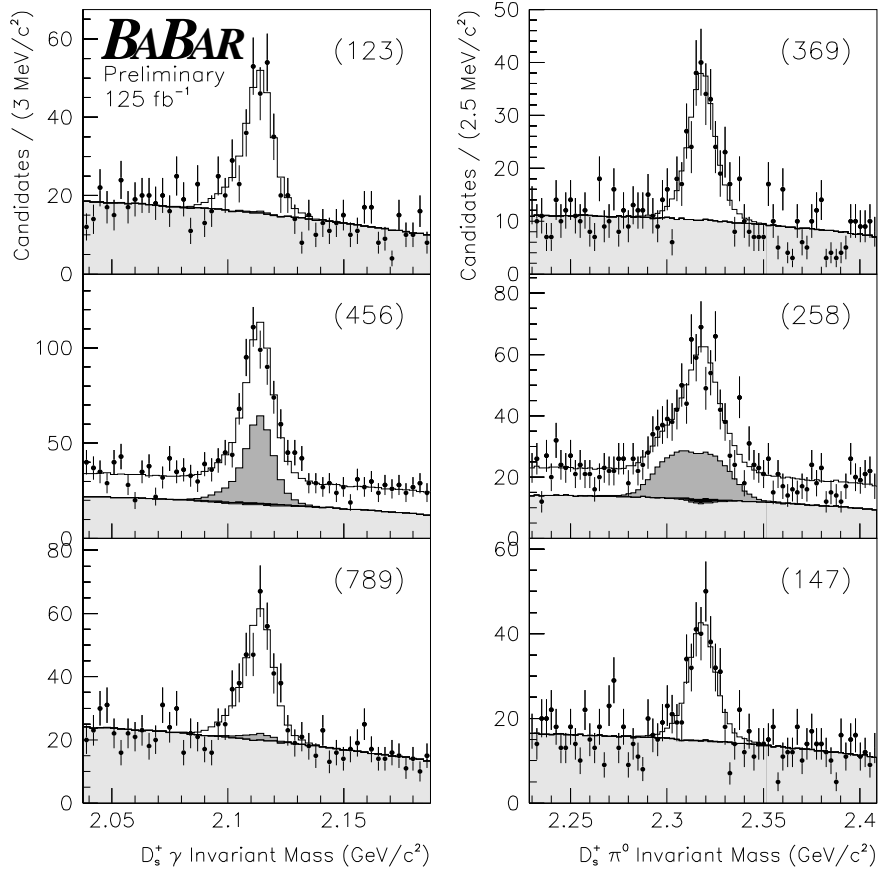}
\end{center}
\vskip -0.35in
\caption{\label{fg:dspi0gam.fit2}
The results of a likelihood fit to the $\Ds\piz$ and $\Ds\gamma$
masses of the $\Ds\piz\gamma$ candidates which appear inside 
the grid shown in
Fig.~\ref{fg:dspi0gam.range}. Each plot corresponds to a column or
row of that grid as indicated by the numbers in parentheses. 
The points correspond to the data.
The histograms are a Monte Carlo
representation of the fit results. Shown in gray (black) are the
contributions from $\DsFE\to\DsTO\piz$
($\DsFE\to\DsTT\gamma$) decay predicted by the fit.
The light gray area is the contribution
from combinatorial background.
}
\end{figure}

The results of this likelihood fit is shown in Fig.~\ref{fg:dspi0gam.fit2},
divided into the regions delineated by the grid shown in
Fig.~\ref{fg:dspi0gam.range}. The fit produces an adequate model of
the data in all regions. The result (statistical errors only)
is a $\DsTO\piz$ yield
of $266\pm 38$ mesons and a $\DsTT\gamma$ yield of $-11\pm 37$ mesons,
the former number being somewhat smaller than the yield determined by
the first fit, though consistent within systematic uncertainties.
Based on these
results, it appears that the decay $\DsFE\to\Ds\piz\gamma$ can be
successfully 
described as proceeding entirely through the channel $\DsTO\piz$.

The likelihood fit of Fig.~\ref{fg:dspi0gam.fit2} is sensitive
to various assumptions. For example, the $\DsTT\gamma$ yield can be
raised to $2$ candidates while preserving the quality of the fit
if the mass resolutions are degraded slightly. In addition,
possible inaccuracies in the inclusive
$\DsTO$ $p^*$ distribution predicted by
Monte Carlo can produce changes in the $\DsTO$ lineshape that
can raise the yield of $\DsTO\piz$ decays to $280$.

\section{\boldmath THE $\Ds\pip\pim$ SYSTEM}
\label{sec:ds2pi}

\begin{figure}
\begin{center}
\includegraphics[width=0.7\linewidth]{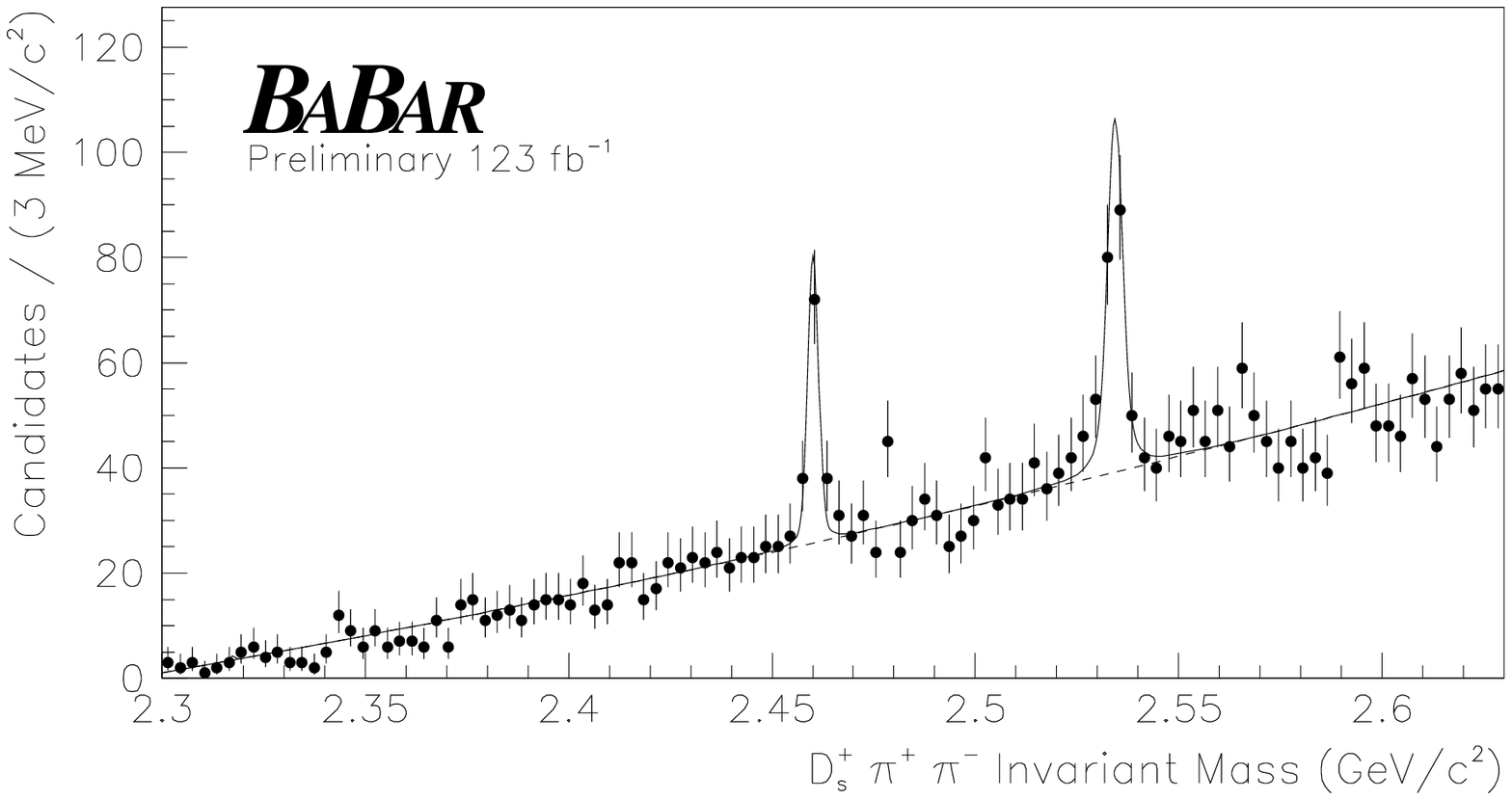}
\end{center}
\vskip -0.35in
\caption{\label{fg:dspipm.fit}The $\Ds\pip\pim$ invariant
mass distribution for candidates that satisfy the requirements
discussed in the text. The solid curve is the result of a unbinned
likelihood fit.
}
\end{figure}

To form $\Ds\pip\pim$ candidates, each $\Ds$ is combined with
$\pip$ and $\pim$ candidates with momentum
above 250~\mevc. This momentum requirement is obtained from the
optimization of $Q$ for the decay $\DsFE\to\Ds\pip\pim$. The
resulting invariant mass distribution has two distinct, narrow
peaks, as shown in Fig.~\ref{fg:dspipm.fit}. These two peaks correspond
to the decays of the $\DsFE$ and $\DsTS$ mesons, as previously
noted by the BELLE collaboration~\cite{Abe:2003jk}.

There is a type of reflection that could be
introduced into the inclusive distribution of $\Ds\pip\pim$ mass
that would not appear in the other combinations discussed up to this point.
An example of such a reflection is the hypothetical decay of
a heavy particle into $\phi$ or $K^*$
plus two charged particles, either of which might be a $K$ meson or proton.
The corresponding $\phi$ or $K^*$ from this decay can be combined
with an unassociated $\pi^+$ candidate to form a fictitious $\Ds$ candidate.
The result could produce a peak in the $\Ds\pip\pim$ mass distribution
when combined with the other decay products.
As a sanity check, the $\Ds$ decay products are
combined with the $\pi^\pm$ candidates using various different
particle species hypotheses ($\pi^\pm$, $p$, $K^\pm$)
to check for underlying resonances
not associated with $\Ds$ decay. No such resonances are uncovered.

The likelihood fit to the $\Ds\pip\pim$ mass distribution consists
of three signal distributions ($\DsTT$, $\DsFE$, or $\DsTS$ decay)
plus a third-order polynomial to describe the combinatorial background.
The shapes of the three signals are derived from signal Monte Carlo
samples. The $\DsFE$ and $\DsTS$ signal shapes are allowed to shift
upwards and downwards in mass in order to best describe the data.
The result of this fit, included in Fig.~\ref{fg:dspipm.fit},
is (statistical errors only)
a $\DsTT$ yield of $0.6\pm 1.8$ decays; a $\DsFE$ mass and yield
of [$2460.1\pm 0.3$]~\mevcc and $67 \pm 11$ decays; 
and a $\DsTS$ mass and yield
of [$2534.3\pm 0.4$]~\mevcc and $124\pm 18$ decays.

A systematic uncertainty specific to this likelihood fit is the
assumption of the background shape. Substituting
a fourth-order polynomial for the background changes the
$\DsFE$ and $\DsTO$ yields by a relative 5.3\% and 9.1\%.

\section{\boldmath THE $\Ds\pi^\pm$ SYSTEM}
\label{sec:dspi}

There has been some conjecture~\cite{Barnes:2003dj,Lipkin:2003zk,Browder:2003fk}
that the $\DsTT$ may be a four-quark hybrid state. It might be expected,
if this was true, that neutral and doubly-charged partners should exist
with a similar mass. The $\Ds\pi^\pm$ system can be used to test this
possibility.

\begin{figure}
\begin{center}
\includegraphics[width=0.7\linewidth]{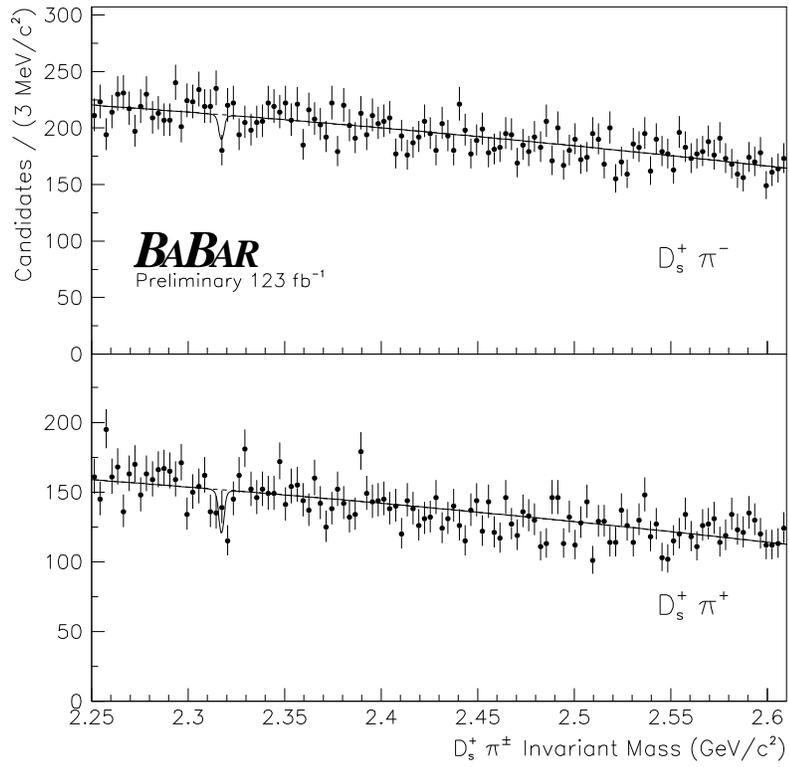}
\end{center}
\vskip -0.35in
\caption{\label{fg:dspi.fit}The $\Ds\pim$ (top) and $\Ds\pip$ (bottom)
invariant mass distributions for candidates that satisfy the requirements
discussed in the text. The solid curve is the result of a unbinned
likelihood fit.
}
\end{figure}

To form $\Ds\pi^\pm$ combinations, each $\Ds$ candidate is combined
with $\pi^\pm$ candidates with momentum greater than 300~\mevc.
This momentum requirement is obtained from the
optimization of $Q$ for the hypothetical decay
$\DsTTz\to\Ds\pim$. The resulting mass distributions are shown
in Fig.~\ref{fg:dspi.fit}. No resonant structure is observed.

Likelihood fits have been performed assuming a hypothetical
$\DsTTz$ or $\DsTTdc$ state at precisely the $\DsTT$ mass.
The results of these fits are included in Fig.~\ref{fg:dspi.fit}.
The yields obtained for the fit (statistical errors only)
for the $\DsTTz$ and $\DsTTdc$
are $-28\pm 25$ and $-39\pm 16$ mesons, respectively.

\section{SYSTEMATIC STUDIES}
\label{sec:Systematics}

There are various systematic uncertainties in common to many of the
likelihood fits presented here. As discussed earlier, there is
an uncertainty of $\pm 0.6$~\mevcc in the given $\Ds$ mass.
A 1\% uncertainty in the energy calibration of the EMC results
in uncertainties of between $0.6$ and $0.8$~\mevcc for those combinations
which include a $\piz$ or $\gamma$ candidate. The largest uncertainty
in the $\pi^\pm$ momentum arises from the accuracy of the calculations 
applied by the track reconstruction software
to account for the energy loss in the material of the detector.
A conservative estimate of the effect on the
$\Ds\pip\pim$ mass of all tracking uncertainties is $\pm 1$~\mevcc.

The mass lineshapes used in the likelihood fits of this analysis,
derived from Monte Carlo simulation, provide models of the data
of satisfactory quality. Variations in the resolution of the detector
have been applied to widen these lineshapes within reasonable bounds.
The resulting change in yield or mass is assigned as a systematic
uncertainty.

As mentioned earlier, the reconstruction efficiency for all particle
combinations discussed in this paper vary as a function of $p^*$.
Yields are calculated by performing separate likelihood fits
with candidates weighted by the inverse of this efficiency. To check
this approach, likelihood fits are also performed on 
the candidate sample divided
into bins of $p^*$. The yields from the binned data are calculated
using the average reconstruction efficiency in that bin. Any difference
in yield produced by these two methods is assigned as a systematic 
uncertainty.

Any inaccuracies in the Monte Carlo predictions of the reconstruction
efficiency will affect the calculation of branching ratios.
As already discussed, the uncertainty in $\piz$ or $\gamma$ efficiency
is assumed to be 3\% and the uncertainty in tracking
efficiency is assumed to be 1.3\%.

\section{COMBINED RESULTS}
\label{sec:Physics}

A preliminary measurement of the $\DsTT$ mass has been obtained from 
the $\Ds\piz$ system:
\begin{equation}
m(\DsTT) = 2318.9\pm 0.3\;({\rm stat.}) 
          \pm 0.9\;({\rm syst.})\;{\rm \mevcc} \;.
\end{equation}

Preliminary measurements of the $\DsFE$ has been obtained from $\Ds\gamma$,
$\Ds\piz\gamma$, and $\Ds\pip\pim$ decays. 
The results, in units of \mevcc, are:
\begin{eqnarray}
m(\DsFE\to\Ds\gamma) &=& 2457.2\pm 1.6\;({\rm stat.}) 
          \pm 1.3\;({\rm syst.}) \\
m(\DsFE\to\Ds\piz\gamma) &=& 2459.1\pm 1.3\;({\rm stat.}) 
          \pm 1.2\;({\rm syst.}) \\
m(\DsFE\to\Ds\pip\pim) &=& 2460.1\pm 0.3\;({\rm stat.}) 
          \pm 1.2\;({\rm syst.}) \;.
\end{eqnarray}
The average of these results, after accounting for correlated
systematic uncertainties, is:
\begin{equation}
m(\DsFE) = 2459.4 \pm 0.3\;({\rm stat.}) 
          \pm 1.0\;({\rm syst.})\;{\rm \mevcc} \;.
\end{equation}

The fit to the $\Ds\pip\pim$ mass spectrum also provides a preliminary
estimate of the $\DsTS$ mass:
\begin{equation}
m(\DsTS) = 2534.3\pm 0.4\;({\rm stat.}) 
          \pm 1.2\;({\rm syst.})\;{\rm \mevcc} \;.
\end{equation}

The following preliminary
branching ratios have been estimated by comparing the
efficiency corrected yields from the corresponding decay modes.
\begin{equation}
\frac{{\mathcal Br}(\DsFE\to\Ds\gamma)}{{\mathcal Br}(\DsFE\to\Ds\piz\gamma)}=
0.375 \pm 0.054\;({\rm stat.}) \pm 0.057\;({\rm syst.})
\end{equation}

\begin{equation}
\frac{{\mathcal Br}(\DsFE\to\Ds\pi^+\pi^-)}{{\mathcal Br}(\DsFE\to\Ds\piz\gamma)}=
0.082 \pm 0.018\;({\rm stat.}) \pm 0.011\;({\rm syst.})
\end{equation}

A 95\% CL upper limit has been calculated for the decay modes in which no 
significant signal is observed. This limit is based on the deviation
from the measured yields of 1.96 standard
deviations, derived from the quadrature sum of statistical and 
systematic uncertainties.

\begin{equation}
\frac{{\mathcal Br}(\DsTT\to\Ds\gamma)}{{\mathcal Br}(\DsTT\to\Ds\piz)}
< 0.17
\end{equation}

\begin{equation}
\frac{{\mathcal Br}(\DsFE\to\Ds\piz)}{{\mathcal Br}(\DsFE\to\Ds\piz\gamma)}
< 0.11
\end{equation}

\begin{equation}
\frac{{\mathcal Br}(\DsFE\to\DsTT\gamma)}{{\mathcal Br}(\DsFE\to\Ds\piz\gamma)}
< 0.23
\label{eq:summary.limit3}
\end{equation}

\begin{equation}
\frac{{\mathcal Br}(\DsTT\to\Ds\pip\pim)}{{\mathcal Br}(\DsTT\to\Ds\piz)}
< 2 \cdot 10^{-3}
\end{equation}

\section{CONCLUSION}
\label{sec:Summary}

Based on 125~${\rm fb}^{-1}$ of $e^+e^-\to c\overline{c}$ data, 
the $\DsTT$ and $\DsFE$ masses are measured to be
[$2318.9\pm 0.3\;({\rm stat.}) 
          \pm 0.9\;({\rm syst.})$]~\mevcc and
[$2459.4\pm 0.3\;({\rm stat.}) 
          \pm 1.0\;({\rm syst.})$]~\mevcc, respectively.
Both values are consistent with previous 
measurements~\cite{Besson:2003cp,Abe:2003jk,Krokovny:2003zq}
with perhaps the exception of the $\DsTT$ mass which is 1.6~\mevcc
higher than our previous result~\cite{Aubert:2003pe}.

Significant signals are observed for $\DsTT\to\Ds\piz$,
$\DsFE\to\Ds\gamma$, $\DsFE\to\Ds\piz\gamma$, and
$\DsFE\to\Ds\pip\pim$. The data is consistent with the
$\DsFE\to\Ds\piz\gamma$ decay proceeding entirely through
the channel $\DsTO\piz$. The $\Ds\pip\pim$ result confirms the decay
first observed by Belle~\cite{Abe:2003jk}, although at a rate
that is somewhat lower.

No significant signal is observed for either the
$\Ds\gamma$ or $\Ds\pip\pim$ decays of the $\DsTT$.
A search for the $\Ds\piz$ decay of the $\DsFE$ also produces
no significant signal. Searches for a neutral or doubly-charged
partner for the $\DsTT$ decaying to $\Ds\pip$ or $\Ds\pim$
also produce no significant signal.

\section*{ACKNOWLEDGMENTS}
\label{sec:Acknowledgments}

% Standard acknowledgments paragraph; must always be included.
We are grateful for the 
extraordinary contributions of our \pep2\ colleagues in
achieving the excellent luminosity and machine conditions
that have made this work possible.
The success of this project also relies critically on the 
expertise and dedication of the computing organizations that 
support \babar.
The collaborating institutions wish to thank 
SLAC for its support and the kind hospitality extended to them. 
This work is supported by the
US Department of Energy
and National Science Foundation, the
Natural Sciences and Engineering Research Council (Canada),
Institute of High Energy Physics (China), the
Commissariat \`a l'Energie Atomique and
Institut National de Physique Nucl\'eaire et de Physique des Particules
(France), the
Bundesministerium f\"ur Bildung und Forschung and
Deutsche Forschungsgemeinschaft
(Germany), the
Istituto Nazionale di Fisica Nucleare (Italy),
the Foundation for Fundamental Research on Matter (The Netherlands),
the Research Council of Norway, the
Ministry of Science and Technology of the Russian Federation, and the
Particle Physics and Astronomy Research Council (United Kingdom). 
Individuals have received support from 
CONACyT (Mexico),
the A. P. Sloan Foundation, 
the Research Corporation,
and the Alexander von Humboldt Foundation.

\bibliography{note1001}% Produces the bibliography via BibTeX.

\end{document}